\documentclass[reprint,amsmath,amssymb,aps,]{revtex4-2}
\usepackage{graphicx}
\usepackage{dcolumn}
\usepackage{bm}
\begin{document}
\title{Bidirectional Mamba state-space model for anomalous diffusion}
\author{Maxime Lavaud$^{1,2}$}
\email{maxime.lavaud@u-bordeaux.fr}
\author{Yosef Shokeeb$^1$}
\author{Juliette Lacherez$^1$}
\author{Yacine Amarouchene$^1$}
\email{yacine.amarouchene@u-bordeaux.fr}
\author{Thomas Salez$^1$}
\email{thomas.salez@cnrs.fr}
\affiliation{$^1$Univ. Bordeaux, CNRS, LOMA, UMR 5798, F-33400, Talence, France.\\$^2$Univ. Bordeaux, CNRS, Bordeaux INP, CBMN, UMR 5248, F-33600, Pessac, France.}
\begin{abstract}
Characterizing anomalous diffusion is crucial in order to understand the evolution of complex stochastic systems,
 from molecular interactions to cellular dynamics. In this work, we characterize the performances regarding such a task of Bi-Mamba, a novel state-space 
 deep-learning architecture articulated with a bidirectional scan mechanism. Our implementation is tested on the AnDi-2 challenge datasets among others. Designed for regression tasks, the Bi-Mamba architecture 
 infers efficiently the effective diffusion coefficient and anomalous exponent from single, short trajectories. As such, our results indicate the potential practical use of the Bi-Mamba architecture 
 for anomalous-diffusion characterization.  
 \end{abstract}
\maketitle

Anomalous diffusion~\cite{Metzler2000} is an ubiquitous phenomenon in complex stochastic processes, and is in particular central to the cell machinery, as classically witnessed by microrheology~\cite{Mackintosh1999}. Its understanding provides insights on the transport of microparticles in heterogeneous media, like biomolecules in biological cells~\cite{Sharifian_2021,Yang2015-vw}. 
Methods such as microscopy 
combined with single-particle tracking allow for the detailed analysis of where and when single events take place \cite{Dahan2003-rh,cognet2007stepwise}. Over the last few decades,
  single-particle imaging methods have been steadily upgraded, increasing the amount of experimental data available on molecular interactions, cellular dynamics, and the behavior of microparticles~\cite{matse2017test,Lavaud2021}, and nanoparticles~\cite{Vilquin2023} in various complex media.
 Other approaches, such as dynamic light scattering \cite{Michael_Schurr1977-rp, Stetefeld2016-ht} and differential dynamic microscopy \cite{Cerbino2017-ib} provide valuable insights too when working with a larger number of objects. 
 
 However, in most practical cases relevant to biological systems, trajectories are typically scarce, short and noisy. As such, it is often hard -- if not impossible -- to infer meaningful information. Deep-learning methods for advanced microscopy have thus emerged as a promising change of paradigm~\cite{deeptrack}. In this context, the work we present here results from the Anomalous Diffusion (AnDi) challenge~\cite{Munoz-Gil2021-dp}, and in particular from our participation to its second edition~\cite{Munoz-Gil2023-mj}. The approach we chose was to evaluate the performances regarding anomalous-diffusion characterization of Bi-Mamba, a novel state-space 
 deep-learning architecture (Mamba) articulated with a bidirectional (Bi) scan mechanism. Indeed, to the best of our knowledge, the Mamba architecture has not been used so far in the case of physical studies, and anomalous 
data in particular. 
This way, we took a step aside from the standard methods used by the other participants, in order to improve the collective knowledge on advanced artificial-intelligence methods for complex particle tracking. Our results indicate that the Bi-Mamba model 
 infers efficiently the effective diffusion coefficient and anomalous exponent from single, short and noisy trajectories.
  
\begin{figure}
    \centering
    \includegraphics[width=1\linewidth]{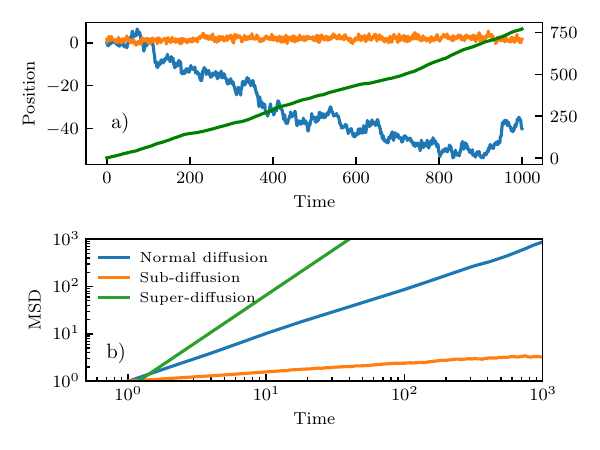}
    \caption{a) Typical trajectories numerically simulated for a Brownian particle ($ \alpha = 1 $), a sub-diffusive particle with $ \alpha = 0.2 $, and a 
    super-diffusive particle with $ \alpha = 1.98. $ b) Mean-squared displacements of the trajectories in a).}
    \label{fig1}
\end{figure}

In a typical two-dimensional anomalous-diffusion process, the mean-squared displacement (MSD) is determined over time $t$ by:
\begin{equation}
   \mathrm{MSD}(t) = 4Kt^\alpha\ ,
\end{equation}
where $ K $ is the effective diffusion coefficient and $ \alpha $ is the anomalous diffusion exponent. For Brownian particles,
  $ \alpha = 1 $. Besides, a particle is in a sub-diffusive regime when  $ 0 < \alpha < 1 $, and in a super-diffusive regime
   when $ 1 < \alpha < 2 $. In Fig.~\ref{fig1}, typical samples of these different types of diffusion are shown with their respective trajectories and MSDs.
In complex cell environments, a particle can exhibit anomalous diffusion due, for example, to:
\begin{itemize}
\item\textit{Obstacles and crowding:} In the presence of obstacles, diffusion can be hindered, 
leading to sub-diffusive behavior. Moderate concentrations of obstacles can cause anomalous diffusion over short distances, and the effect becomes more pronounced as the obstacle concentration approaches the percolation threshold \cite{Saxton1994Anomalous};
    
\item\textit{Binding and trapping:} Temporary binding of particles to fixed sites~\cite{Saxton1996AnomalousDD} 
leads to a sub-diffusive behavior;
    
\item\textit{Lipid rafts and membrane heterogeneity:} Interactions of particles with lipid rafts in cell membranes have been shown to lead to complex diffusive behaviors 
\cite{Nicolau2007Sources};
    
\item\textit{Chaos:} Deterministic chaos and intermittency can also lead to anomalous diffusion, as shown in \cite{Geisel1984Anomalous}. 
Deterministic systems can produce a non-linear growth of the MSD due to long-time correlations and chaotic mechanisms.
\end{itemize}

The first AnDi Challenge took place in 2020 and aimed at assessing  the performances of various methods in quantifying anomalous diffusion~\cite{Munoz-Gil2021-dp}. The main focus was dedicated to the inference of $K$ and $\alpha$, for various simulations of bio-mimetic cases. The second AnDi challenge took place in 2024 and aimed at evaluating the performances of various methods for detecting and quantifying changes in single-particle motion~\cite{Munoz-Gil2023-mj}. As such, the focus was not only on the inference of the effective diffusion coefficient and exponent, but also on trajectory segmentation, with the trajectory statistics being described by a maximum of two $K$ and two $\alpha$ values. One typical goal is for example to measure at what rate a particle binds and unbinds to/from a given cellular site. 
The AnDi datasets involve five phenomenological models:
\begin{itemize}
\item\textit{Single-state:} particles diffusing according to a single diffusion state, as observed for some lipids in the plasma membrane~\cite{Eggeling2009-wn, Manzo2011-zm, Honigmann2014-uq};

\item\textit{Multi-state:} particles diffusing according to two diffusion states, and undergoing transient changes of $K$ and $\alpha$, 
as observed for proteins due to allosteric changes or ligand binding~\cite{Mainali2013-ji,Yanagawa2018-cd,Da_Rocha-Azevedo2020-rn,Achimovich2023-oi};

\item\textit{Dimerization:} particles diffusion according to two diffusion states, and undergoing transient changes of $K$ and $\alpha$, induced by encountering 
other diffusing particles, as observed in protein dimerization and protein-protein interactions~\cite{Low-Nam2011-hh, Grimes2022-yl};

\item\textit{Transient confinement:} particles undergoing transient diffusion changes when entering or 
leaving given areas, as observed in the confinement induced by clathrin-coated pits on cell membranes~\cite{Weigel2013-ls};

\item\textit{Quenched trap:} two-state model of diffusion 
representing proteins being transiently immobilized at specific locations, induced by binding to immobile structures, as observed  in cytoskeleton-induced molecular pinning~\cite{Spillane2014-lx, Rossier2012-rl}.
\end{itemize}

Deep-learning models for time-series inference are generally based on Recurrent Neural Networks (RNNs) or Convolutional Neural Networks (CNNs). 
RNNs process the data sequence by updating a hidden state after each element. However, as the gradient also needs to go through 
each element one by one, long-term information may be lost. CNNs use convolutional kernels to give more weight to local information,
 as they do not need to process the sequence one by one. CNNs benefit from parallel computing and faster training speed but also limit the retrieval of 
 long-term global information. As a consequence, the Mamba model~\cite{Gu2023MambaLS}, a state-space model, has been recently developed to overcome the long-term 
 loss of information while being fast to train. The Mamba architecture is essentially a reformulation of RNNs and CNNs as selective state-space models. The novelties brought by this architecture consist in:
\begin{itemize}
\item\textit{A selection mechanism:} allows the model to ignore irrelevant information or focus on relevant information in an input-dependent manner, 
which is comparable to the attention mechanism in Transformers;
\item\textit{A hardware-awareness:} the model is optimized for use on a GPU, allowing it to scale linearly with sequence length.
\end{itemize}
\begin{figure}
    \centering
    \includegraphics[width=1\linewidth]{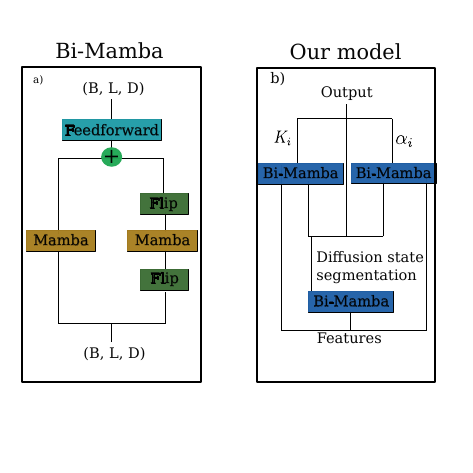}
    \caption{a) Bi-Mamba block implementation. b) Deeper Bi-Mamba implementation.}
    \label{fig:fig2}
\end{figure}

By drawing inspiration from Bi-Mamba+ \cite{liang2024bimambabidirectionalmambatime},
 we constructed a deep-learning model based on the Mamba architecture with a bidirectional scan mechanism. As shown in Fig.~\ref{fig:fig2}a), 
 this scheme involves using the features and their time-flipped counterparts within two different Mamba blocks. The outputs are then concatenated and sent 
 to a feedforward layer. Such a bidirectional approach ensures that the time dependencies across the entire trajectory are captured more effectively. As shown in Fig.~\ref{fig:fig2}b),  our model architecture consists instead of three blocks. First, the trajectories are passed 
into a Bi-Mamba block for segmentation, where the diffusion modes are hot-encoded. Then, the trajectories and segmentation results are forwarded to the next 
Mamba blocks, one dedicated to the $K$ regression and the other to the $\alpha$ regression. Therefore, each task is handled by a dedicated block,
 allowing each block to specialize in a single function. Segmentation is evaluated using the Weighted Cross-Entropy Error (WCE) loss, while the $K$ regression is evaluated using the Mean-Squared 
 Logarithmic Error (MSLE) loss, and the $\alpha$ regression is evaluated using the Mean Absolute Error (MAE) loss. The losses are then summed, and the total loss is back-propagated using the Adam optimizer.

\begin{figure}
    \centering
    \includegraphics{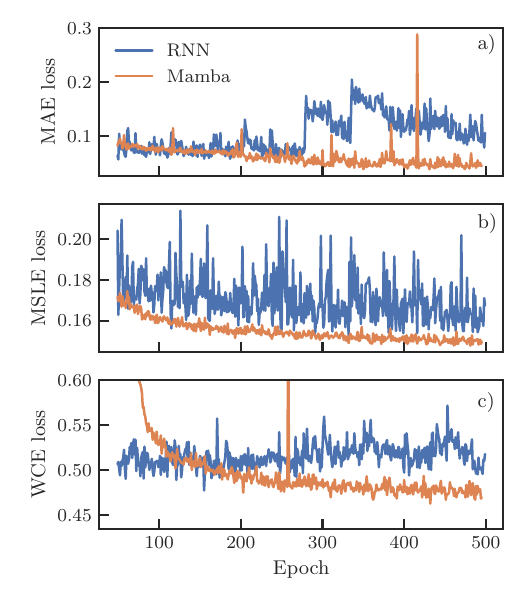}
    \caption{Loss values calculated on a test dataset consisting of $10^4$ trajectories, as functions of the number of epochs. Specifically, we show: a) 
    the Mean Absolute Error (MAE) loss for $\alpha$ inference, b) the Mean-Squared Logarithmic Error (MSLE) loss for $K$ inference, and c) the Weighted Cross-Entropy (WCE) loss for diffusion-state segmentation.}
    \label{fig:mambavsothers}
\end{figure}

The data ensemble consists of sets of two-dimensional trajectories with coordinates $r(t) = [r_x(t), r_y(t)]$, with a maximum length of 200 time steps. 
The features computed using given trajectories are: the displacements for the first lag time $\Delta t = 1 $, $\Delta r_i (\Delta t =1) = r_i(t+ \Delta t) - r_i(t )$, 
with $i \in \{x, y\}$, the one-dimensional MSDs $\langle \Delta r_i ^2\rangle (\Delta t) $, the displacement angle between two consecutive displacements
 \cite{Kabbech2024}, and the total displacement from the origin $d = \sqrt{ x(t)^2  + y(t)^2}$. Trajectories smaller than 200 data points are zero-padded. For fair benchmarking purposes, we compare our model to a bidirectional RNN, as shown in Fig.~\ref{fig:mambavsothers}. The Bi-Mamba model shows promising performances,
scoring better overall and in each category of the AnDi2 Challenge. Moreover, it shows its capabilities to train for a larger number of epochs without overfitting and with a smaller loss variance.

In this work, we implemented the structure and evaluated the performances of the Mamba architecture -- a novel application of state-space models -- towards the characterization of anomalous diffusion. Using further a bidirectional scan 
mechanism, we demonstrated a notable efficiency in diffusion-state segmentation, as well as for effective-diffusion-coefficient and anomalous-exponent regression tasks. Improvement could be achieved through the use of specialized models,
 fine-tuned using simulations generated near the predictions of the generally-trained model presented in this work.
Our contribution to the AnDi-2 challenge was designed in order to specifically evaluate the performances of Bi-Mamba models for anomalous diffusion. Our latest model ranked 7th in $\alpha$ inference (MAE of $0.27$), 9th in $K$ inference (MSLE of $0.05$), 3rd in diffusion-type measurement [\textit{i.e.} trapped with ($\alpha < 0.2$), directed with $\alpha > 1.8$, or normal with $\alpha =1$ and with a F1 error of $0.91$], and 10th on the change-point detection (Root-Mean-Squared Error (RMSE) of $2.7$). Moreover, we expect improvements on the segmentation task by using Mamba-based Unet methods~\cite{U-Mamba}. All together, this work contributes to the knowledge on, and continued development of, deep-learning methods for anomalous diffusion and physics in general.  

\section*{Data availability}
Our Bi-Mamba implementation for anomalous diffusion is available at
https://github.com/EMetBrown-Lab/Mamba-EMetBrown-ANDI2.

\section*{Acknowledgements} 
Computer time for this study was provided by the computing facilities of the M\'esocentre de Calcul Intensif Aquitain.
 The authors acknowledge financial support from the European Union through the European Research Council under EMetBrown 
 (ERC-CoG-101039103) grant. Views and opinions expressed are however those of the authors only and do not necessarily reflect 
 those of the European Union or the European Research Council. Neither the European Union nor the granting authority can be held 
 responsible for them. The authors also acknowledge financial support from the Agence Nationale de la Recherche under
  Softer (ANR21-CE06-0029) and Fricolas (ANR-21-CE06-0039) grants, as well as from the Interdisciplinary and 
 Exploratory Research Program under a MISTIC grant at the University of Bordeaux, France. They acknowledge as well the support from the LIGHT
 S{\&}T Graduate Program (PIA3 Investment for the Future Program, ANR-17EURE-0027). Finally, they thank the RRI Frontiers of Life, which 
  received financial support from the French government in the framework of the University of Bordeaux's France 2030 program,
   as well as the Soft Matter Collaborative Research Unit, Frontier Research Center for Advanced Material and Life Science, 
 Faculty of Advanced Life Science, Hokkaido University, Sapporo, Japan, and the CNRS International Research Network between France and India on ``Hydrodynamics at small scales: from soft matter to bioengineering".
\bibliographystyle{apsrev4-2}
\bibliography{Lavaud2024}
\end{document}